\documentclass{Interspeech2024}
\usepackage{multirow} 
\usepackage{tabularx}
\usepackage{subfig}
\usepackage{graphicx}
\usepackage{caption}
\usepackage[symbol]{footmisc}
\usepackage[normalem]{ulem}

\interspeechcameraready

\title{Analyzing Speech Motor Movement using Surface Electromyography \\ in Minimally Verbal Adults with Autism Spectrum Disorder}

\name[affiliation={1}]{Wazeer}{Zulfikar$^*$}
\name[affiliation={1}]{Nishat}{Protyasha$^*$}
\name[affiliation={2}]{Camila}{Canales}
\name[affiliation={2}]{Heli}{Patel}
\name[affiliation={3}]{James}{Williamson}
\name[affiliation={2}]{Laura}{Sarnie}
\name[affiliation={2}]{Lisa}{Nowinski}
\name[affiliation={1}]{Nataliya}{Kosmyna}
\name[affiliation={2, 4}]{Paige}{Townsend}
\name[affiliation={3}]{Sophia}{Yuditskaya}
\name[affiliation={3}]{Tanya}{Talkar}
\name[affiliation={1}]{Utkarsh Oggy}{Sarawgi$^*{^*}$}
\name[affiliation={2}]{Christopher}{McDougle}
\name[affiliation={3}]{Thomas}{Quatieri}
\name[affiliation={1}]{Pattie}{Maes}
\name[affiliation={2, 4}]{Maria}{Mody}

\address{
  $^1$MIT Media Lab, USA
  $^2$MGH Lurie Center for Autism, USA \\
  $^3$MIT Lincoln Lab, USA 
  $^4$MGH Martinos Center for Biomedical Imaging, USA}
\email{\{wazeer, protya\}@mit.edu}


\begin{document}

\maketitle

\renewcommand{\thefootnote}{\fnsymbol{footnote}}
\footnotetext[1]{Equal Contribution}
\footnotetext[7]{Work done while at MIT Media Lab, currently at Apple}

\begin{abstract}

Adults who are minimally verbal with autism spectrum disorder (mvASD) have pronounced speech difficulties linked to impaired motor skills. Existing research and clinical assessments primarily use indirect methods such as standardized tests, video-based facial features, and handwriting tasks, which may not directly target speech-related motor skills. In this study, we measure activity from eight facial muscles associated with speech using surface electromyography (sEMG), during carefully designed tasks. The findings reveal a higher power in the sEMG signals and a significantly greater correlation between the sEMG channels in mvASD adults (N=12) compared to age and gender-matched neurotypical controls (N=14). This suggests stronger muscle activation and greater synchrony in the discharge patterns of motor units. Further, eigenvalues derived from correlation matrices indicate lower complexity in muscle coordination in mvASD, implying fewer degrees of freedom in motor control.
\end{abstract}

\section{Introduction}

Approximately 30\% of adults with autism spectrum disorder (ASD) are estimated to be minimally verbal (MV) \cite{maltman2021brief}. Alongside shared ASD traits, minimally verbal autism spectrum disorder (mvASD) is distinguished by severe difficulty with speech and is associated with reduced expressive language skills. A vast majority of the ASD population, about $79\%$ to $88\%$, have motor impairments \cite{licari2019, bhat2021} and the development of language in ASD is well-known to be influenced by such impairments \cite{maffei2023oromotor}. Prior research suggests that motor coordination and control difficulties, particularly in facial and fine motor gestures, are likely to contribute to speech differences in ASD \cite{maffei2023oromotor, mccleery2013motor}. 

Despite the known significance of reduced oral motor skills in ASD speech challenges \cite{maffei2023oromotor}, the underlying physiological markers remain underexplored. Previous works have explored motor skills and coordination in ASD primarily through perceptual studies and acoustic analysis \cite{talkar2020assessment, mody2017communication, chenausky2019motor, finemotor}.

In this paper, we present a physiological examination of motor patterns associated with speech production in mvASD by analyzing neuromuscular action potential signals \cite{farina2004extraction}. We utilize surface electromyography (sEMG) to quantify movements of facial muscles in adults with mvASD, comparing these movements with those of age- and gender-matched neurotypical adults during speech tasks. These tasks are tailored to the communication limitations in mvASD and include the clinically used diadochokinesis sequence \cite{ziegler2002task} and a select set of six words elicited under varying cognitive demands.

We present preliminary results comparing sEMG signals from facial muscles associated with speech production in the mvASD and the neurotypical control (NT) groups. Compared to NT, individuals with mvASD exhibit:
\begin{itemize}
\item Higher power of neuromuscular action potentials
\item Significantly greater correlation between the eight sEMG channels
\item Lower complexity in the coordination of facial muscle movements
\end{itemize}
Taken together, the results suggest tightly coupled and less complex
muscle movements with fewer degrees of freedom, and effortful
speech production; evident in the correlation between the sEMG signals
from different channels and higher power of action potentials
from the facial muscles during the speech tasks.

\section{Related Work}

\subsection{Motor Skills and Speech Challenges in ASD}

Orofacial activity has been correlated with speech production \cite{mcclean2003association}. Extensive research on oromotor skills in the ASD population has employed perceptual approaches such as audio/visual assessments \cite{maffei2023oromotor}. However, perceptual analyses are vulnerable to rater's bias \cite{kent1996hearing, chenausky2023review}. Instrumental techniques that provide objective data and reduce perceptual biases include facial motion tracking \cite{talkar2020assessment}, acoustic analysis \cite{patel2020acoustic, sarawgi2020uncertainty, sarawgi2021uncertainty}, ultrasound imaging \cite{mckeever2022using}, magnetoencephalography (MEG) \cite{pang2016abnormal}, and electromyography (EMG) \cite{pascolo2012relationship}. sEMG is a non-invasive technique for directly examining speech-motor physiology by recording electrical activity, known as action potentials, generated by muscle cells. Kapur et al \cite{alterego} used sEMG from the facial muscles for speech recognition showing its capabilities to capture critical oromotor movements associated with speech production. Limited studies have examined oromotor muscle activation patterns in individuals with ASD using sEMG \cite{maffei2023oromotor}. However, they have solely been used to study non-speech activity \cite{pascolo2012relationship, cattaneo2007impairment}. To our knowledge, our work is the first to investigate sEMG-based physiological patterns focused on speech production among minimally verbal individuals with ASD.




\subsection{Motor Coordination in Speech Production}

Speech production involves the coordinated effort of muscle groups, characterized by the articulatory, respiratory, and laryngeal subsystems. Cross-correlation analysis of sEMG recordings from multiple muscle groups indicates the degree of synchronization \cite{kilner2002novel}, with strong correlations suggesting similar motor unit discharge patterns \cite{lowery2003simulation}. Previous studies have used correlation structures \cite{williamson2019tracking} derived from acoustic analysis of formant values and facial features as a proxy measure of motor coordination \cite{talkar2020assessment}. These features have been used to predict clinical conditions including characteristics of the motor coordination during speech production in ASD \cite{talkar2020assessment}, and clinical severity scores of individuals with Parkinson's disease \cite{parkinson}, major depressive disorder \cite{williamson2019tracking}, and mild traumatic brain injury \cite{helfer14_interspeech}. In this work, we apply the correlation structures \cite{williamson2019tracking} to the sEMG signals from different facial muscles to characterize the synchronicity and complexity of the motor coordination pattern associated with speech production.

\section{Methods and Materials}
The methods and materials used for the study were approved by Human Subjects Research committees of the hospital and university involved.

\subsection{Participants}
$26$ participants $-$  $12$ with minimally verbal ASD ($8$ male, $4$ female) and $14$ neurotypical controls ($8$ male, $6$ female) $-$ between the ages of $19-32$ years ($23.85\pm3.99$ years) were recruited from MGH Lurie Center for Autism. There was no significant difference in age between the groups ($p>0.05$). The diagnosis of ASD was confirmed by an experienced clinician using the DSM$-5$ criteria and the Social Communication Questionnaire (SCQ) (American Psychiatric Association, 2013). All participants had normal, or corrected to normal hearing and vision, and English as the primary language. Participants with excessive or abnormal motor movements, medical, neurological, substance use disorders, other comorbid developmental disorders, or known voice pathologies were excluded. NT participants were also excluded if they were medicated for a psychiatric condition or had a family history of ASD in a first-degree family member.

Clinical profiles were evaluated during a screening visit using a battery of standardized neuropsychological assessments designed to measure cognitive, language, and motor skills, including the Stanford Binet$-5$ (SB$-5$), Peabody Picture Vocabulary Test$-5$ (PPVT$-5$), Expressive Vocabulary Test$-3$ (EVT$-3$), Grooved Pegboard task, and Beery Visuomotor Integration task (VMI$-6$). Social skills were also evaluated using the Social Responsiveness Scale$-2$ (SRS$-2$). ASD participants were considered minimally verbal (MV) if they had moderate-to-significant speech impairment based on the Clinical Global Impression-Severity (CGI-S) scale and had expressive language skills $<1^{\text{st}}$ percentile on the EVT. 

\subsection{Protocol}
All participants underwent standardized clinical testing to qualify them for participation in the study. This was followed by an experimental visit during which data was collected on a series of speech production tasks. The tasks were diadochokinetic (DDK) sequencing and elicitation of six selected words. For the DDK task, which is clinically used for diagnosing Apraxia of speech \cite{ziegler2002task}, participants were instructed to repeat the sequences of `pa-pa-pa', `ta-ta-ta', `ka-ka-ka', and `pa-ta-ka', as many times as possible in a single breath. Following that, in the word production task, six words $-$ `book', `dog', `girl', `apple', `kitten', and `potato' $-$ were used. The words were selected based on language acquisition literature \cite{Fenson2023} to cover a range of phonetic features and syllable lengths. The same set of words was used under three different speech production scenarios $-$ imitation, naming, and reading $-$ to incorporate varying cognitive demands; each word was attempted three times in each condition. The tasks were presented on a laptop through the Psychtoolbox software \cite{brainard1997psychophysics}. All recordings were done using a custom-built application in MATLAB to collect simultaneous audio and video data. During the imitation task, participants heard and repeated the word as spoken by the experimenter.  During the naming task, pictures were used for the target words. During the reading task, participants were asked to read aloud printed words and point to the corresponding picture from two choices. Video recordings of the session were used to confirm surface electrode positioning for sEMG and status throughout the recording session; and for any additional facial movement cues relating to initiation of speech-related movements.





\begin{figure}
    \subfloat[Diadochokinetic (DDK) sequencing task]{
    \label{fig:data_ddk}
    \includegraphics[width=\linewidth]{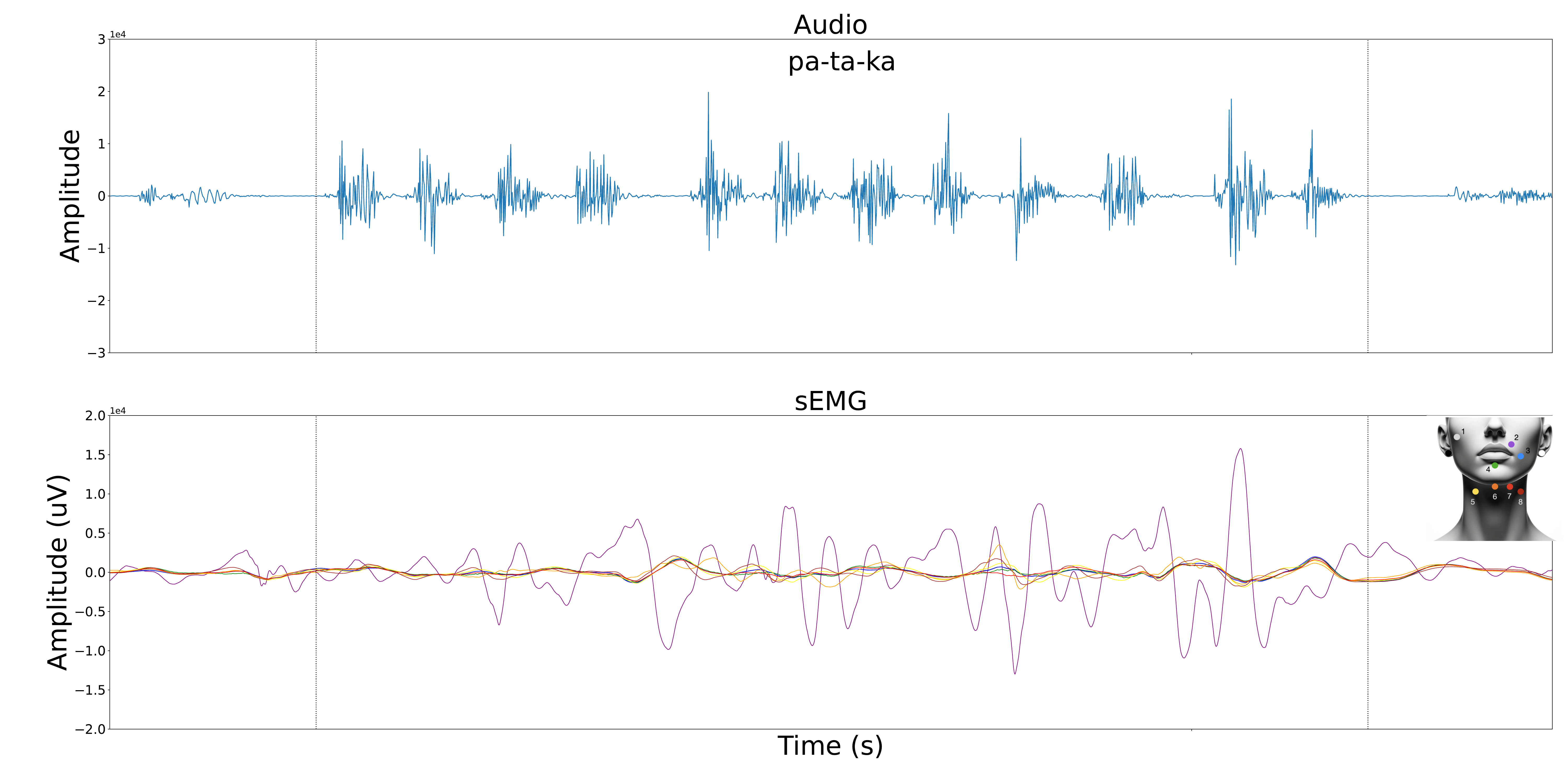}}
    

    \subfloat[Word Production Task]{\label{fig:data_words}    \includegraphics[width=\linewidth]{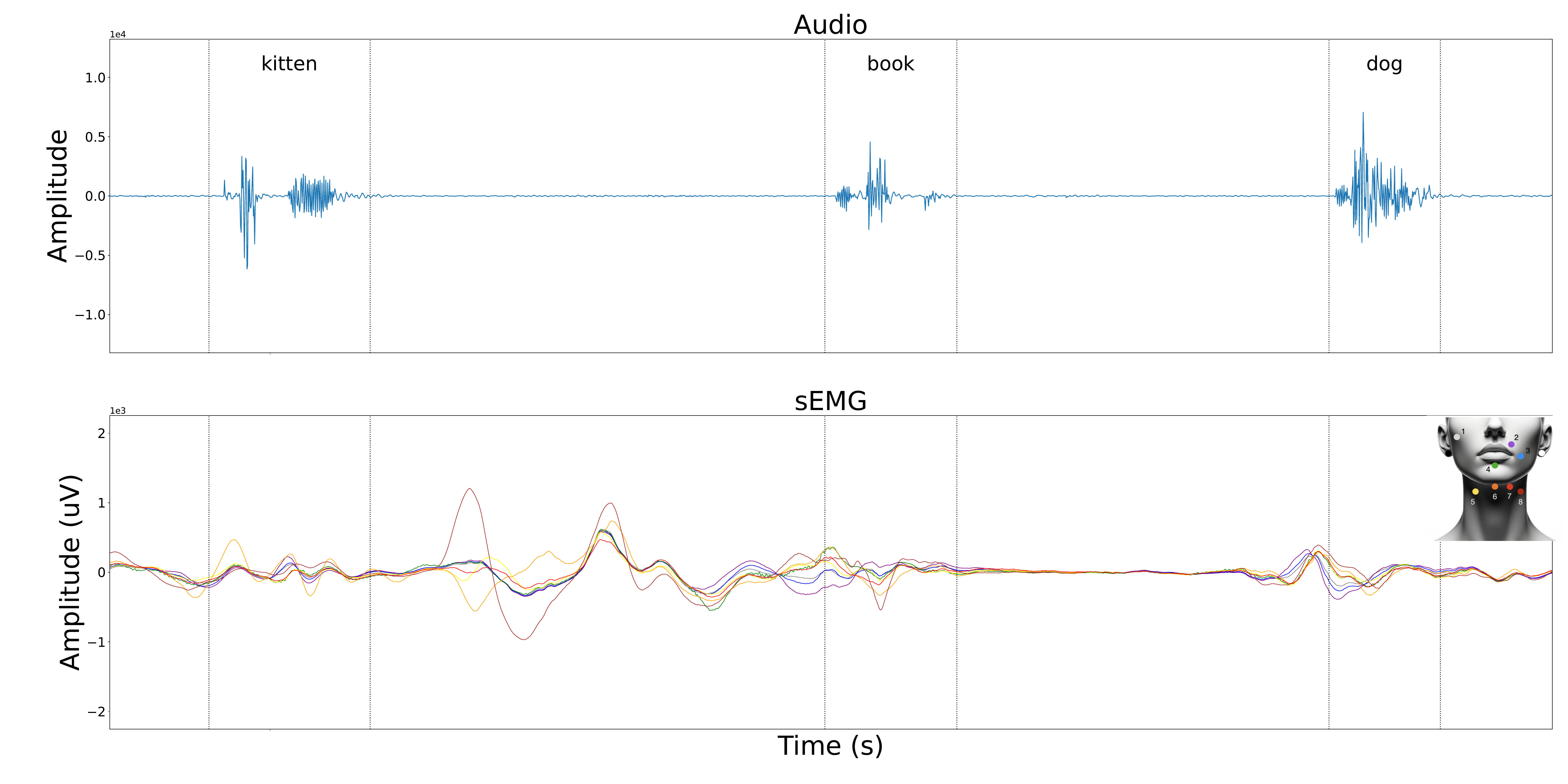}}
    
    \caption{Simultaneous audio and sEMG data collected from an mvASD participant during speech production tasks revealing approximate synchrony across many channels. The different sEMG channels are color-coded.}
\end{figure}

\subsection{sEMG Data Collection}

\begin{figure}[th!]
    \centering
    \includegraphics[width=0.6\linewidth]{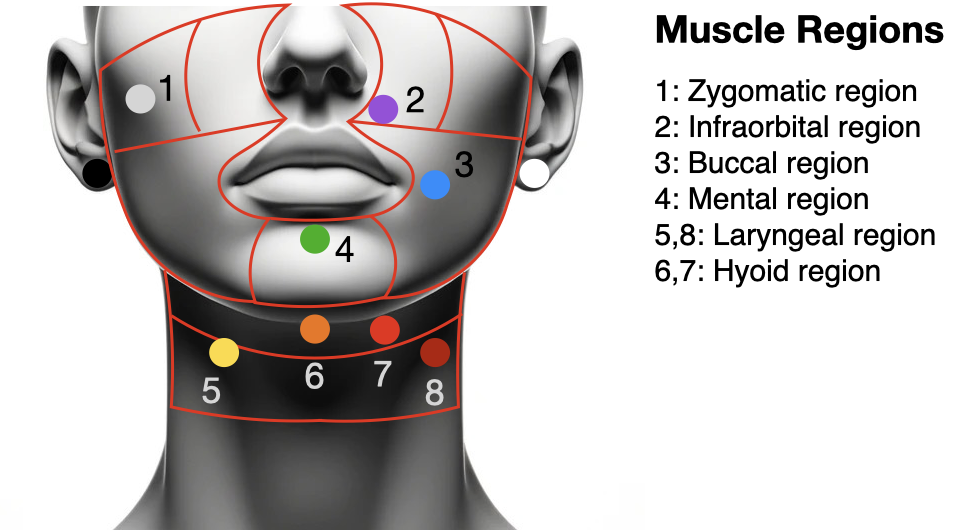}
    \caption{Placement of eight electrodes for surface electromyography (sEMG) with two reference electrodes on the ears.}
    \label{fig:electrode_placement}
    \vspace{-2mm}
\end{figure}

Previous studies have proposed and used a wearable sEMG-based system for measuring neuromuscular activity, i.e. sEMG data from the face and neck regions, to help facilitate limited speech recognition \cite{kapur2020non, alterego}. We followed the same electrode placement (shown in Figure \ref{fig:electrode_placement}), data collection, and processing standards. Non-invasive neuromuscular signals sampled at $250$ \si{\hertz} were recorded using eight channels against reference and bias Ag/AgCl surface electrodes placed on the participant's face and neck, as detailed and illustrated in \cite{kapur2020non}. This setup helps record subtle disruptions in the electric field at facial regions associated with speech production. A high-resolution $24-$bit analog-to-digital converter was used with a graphical user interface for real-time data processing, visual feedback, and annotation based on protocol guidelines. Raw neuromuscular recordings were corrected for baseline drift and DC offset, heartbeat artifacts, and power line noise \cite{kapur2020non}. Finally, the signals were rectified and filtered for frequencies ranging from $0.5$ to $20$ \si{\hertz}. The range was decided to capture movements within and across phonemes based on Hermansky’s RASTA peaks of $8$ \si{\hertz} and cut off at $20$ \si{\hertz} \cite{hermansky1994rasta}. Simultaneous audio and video as collected per protocol and annotation guidelines helped correct for any erratic data. Figure \ref{fig:data_ddk} and Figure \ref{fig:data_words} show respective examples of the data collected for a mvASD participant for the two tasks $-$ the DDK task and the word production task.
 
\subsection{Feature Extraction}
Speech data from the participants were manually segmented into each word utterance and DDK sequence by trained annotators using the open-source Praat software \cite{boersma2001speak}. 

\textit{Power:}  Root mean square (RMS) was computed to estimate the power of each sEMG channel. The RMS values were averaged for all channels for each utterance, and then for each participant across the 9 utterances ($3$ tasks x $3$ attempts). 
Then, the mean RMS for each word and DDK sequence was calculated across all participants within a group and compared between the groups. To test for significance, we used a random mixed effects model with Gaussian distribution to account for the inherent variability across subjects and words/sequences. The test was done across all utterances for the words and sequences combined (subject and words/sequences considered as random effects), and for each word and sequence separately (only subject as random effects). The tests were conducted using the glmmTMB package in R.

\textit{Correlation:} Pearson correlation coefficients were calculated (with zero phase delay) between each of the eight sEMG channels, resulting in a correlation matrix of $8$x$8$ values. The mean of the correlation matrix was determined and averaged over each word independently to perform a group analysis. To test for significance between the mvASD and the NT groups, we used a mixed-effects model similar to what was used to compare power. The Gaussian model was switched to a beta regression model due to the boundedness of the correlation values.

\textit{Correlation Structures:} Multivariate auto- and cross-correlations of the eight sEMG channels were used to produce measures of coordination within and across the facial muscles \cite{williamson2019tracking, parkinson, helfer14_interspeech}. Time-delay embedding was used to expand the dimensionality of the signals, resulting in correlation matrices that represent coupling strengths across channels at multiple relative time delays ($15$ delays of $10$ms each, resulting in a $120$x$120$ matrix). Eigenvalues of the correlation matrices were computed by rank order. Higher eigenvalues in the first few eigenvalue indices and a steeper decline in the lower eigenvalue indices indicate a lower complexity in the coordination of the muscle movements (i.e., fewer degrees of freedom in motor dynamics). In contrast, a more uniform set of eigenvalues across the eigenspectra implies a higher complexity of coordination.

\begin{table*}[h]
\centering
\caption{Mean Root Mean Square (RMS) and mean correlation of sEMG signals from the eight channels across different words/sequences and grouped by participant group. P-values in both features were calculated using mixed-effects models.
}
\begin{tabularx}{0.85\textwidth}{l|c|c|c|c|c|c}
\toprule
\multirow{2.5}{*}{\textbf{Word/ Sequence}} & \multicolumn{3}{c}{\textbf{Root Mean Square (\si{\micro\volt})}} &   \multicolumn{3}{c}{\textbf{Correlation}} \\
\cmidrule{2-4} \cmidrule(l){5-7}
& mvASD & NT & p-value & mvASD & NT & p-value  \\
\midrule

\textit{dog} &  $137.6 \pm 34.1$ &  $89.1 \pm 22.0$ & $0.1080$  &  $0.60 \pm 0.06$  & $0.42 \pm 0.04$  & $0.0106$  \\
\hline 
\textit{potato} & $191.1 \pm 64.0$ &  $159.3 \pm 88.6$ & $0.7698 $ &  $0.61 \pm 0.06$ &  $0.43 \pm 0.04$ & $0.0084 $\\
 
\hline 
\textit{kitten} &   $220.9 \pm 83.7$ & $104.5\pm 49.6$ & $0.1452$ &   $0.62 \pm 0.06$ &   $0.48 \pm 0.05$ &  $0.0274 $  \\
 
\hline 
\textit{girl} & $166.0 \pm 56.6$ &  $76.5 \pm 28.3$ & $0.0864$ &   $0.62 \pm 0.06$  &   $0.41 \pm 0.05$ &  $0.0114$ \\
 
\hline 
\textit{book} & $164.9 \pm 51.8$ & $99.6 \pm 39.6$ & $0.1690$ &   $0.56 \pm 0.05$ &  $0.43 \pm 0.03$  & $0.0210$ \\
 
\hline 
\textit{apple} &  $204.2 \pm 65.4$ &  $170.7 \pm 68.4$ & $0.5260$ &   $0.59 \pm 0.06$  &   $0.45 \pm 0.04$ &  $0.0490$\\
 
 \hline 
\textit{pa-pa-pa} &  $178.1 \pm 82.8$ &  $113.9 \pm 39.9$ & $0.4627 $ &  \ $0.64 \pm 0.05$ &   $0.40 \pm 0.05$  &  $0.0016$\\

\hline 
\textit{ta-ta-ta} & $153.0 \pm 55.7$ &  $88.7 \pm 29.7$ & $ 0.2860 $ &   $0.67 \pm 0.06$ &   $0.47 \pm 0.05$ &  $0.0011$\\

\hline 
\textit{ka-ka-ka} &  $140.9 \pm 34.8$ &  $173.8 \pm 107.1$ & $0.9164$ &   $0.71\pm 0.05$ &  $0.50 \pm 0.04$ &  $0.0006$\\

\hline 
\textit{pa-ta-ka} &  $200.5 \pm 91.4$ &  $288.5 \pm 188.3$ & $0.9287 $ & $0.60 \pm 0.06$ &   $0.37 \pm 0.04$ &  $0.0017$  \\

\bottomrule 
\end{tabularx}
\vspace{2mm}

\label{tab:features} 
\end{table*}

\section{Results}

\subsection{Clinical Assessments}

Cognitive ($107.29\pm10.31$), language (receptive: $110.77\pm13.59$; expressive: $111.31\pm9.66$), motor (fine motor: $103.79\pm15.49$; visuomotor integration: $98.92\pm7.04$), and social skills ($49.00\pm9.31$) in neurotypical NTs fell consistently within the average range (defined as cognitive, language, and motor skills between $85-115$, and social skills $<59$). \\
In contrast, mvASD participants showed significantly lower cognitive ($55.25\pm15.98$), language (receptive: $49.08\pm23.66$; expressive: $55.25\pm9.01$), motor (fine motor: $51.00\pm12.93$; visuomotor integration: $49.50\pm13.39$), and social skills compared to NT ($72.42\pm10.48$; $p < 0.001$) with scores consistently in the very low range ($5-6$ year developmental age level).

\subsection{Power of Neuro-muscular Action Potentials}
Table \ref{tab:features} reports the channel-averaged root mean square (RMS) of the preprocessed sEMG signals separately for the mvASD and NT groups. The average RMS values were greater for the mvASD group compared to the NT group for every word and DDK sequence except `ka-ka-ka' and `pa-ta-ka'. However, the significance test indicated that these differences were not statistically significant ($p>0.05$).

\begin{figure}
    \subfloat[Correlation matrix across the eight sEMG channels during diadochokinetic (DDK) task.]{\label{fig:correlation_matrix}    \includegraphics[width=\linewidth]{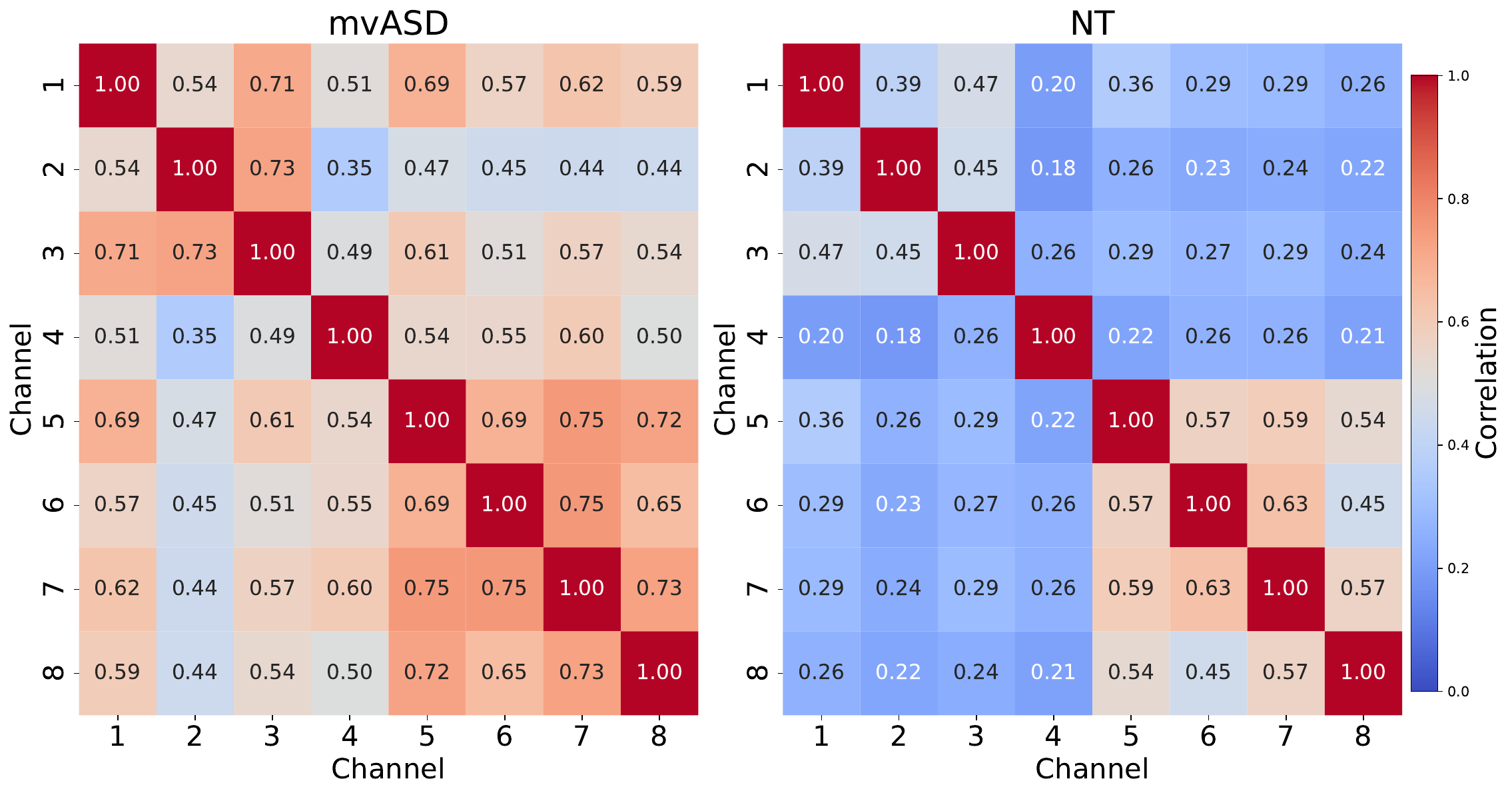}}

    \subfloat[Mean correlation values per group aggregated per word.]{
    \label{fig:correlation}
    \includegraphics[width=\linewidth]{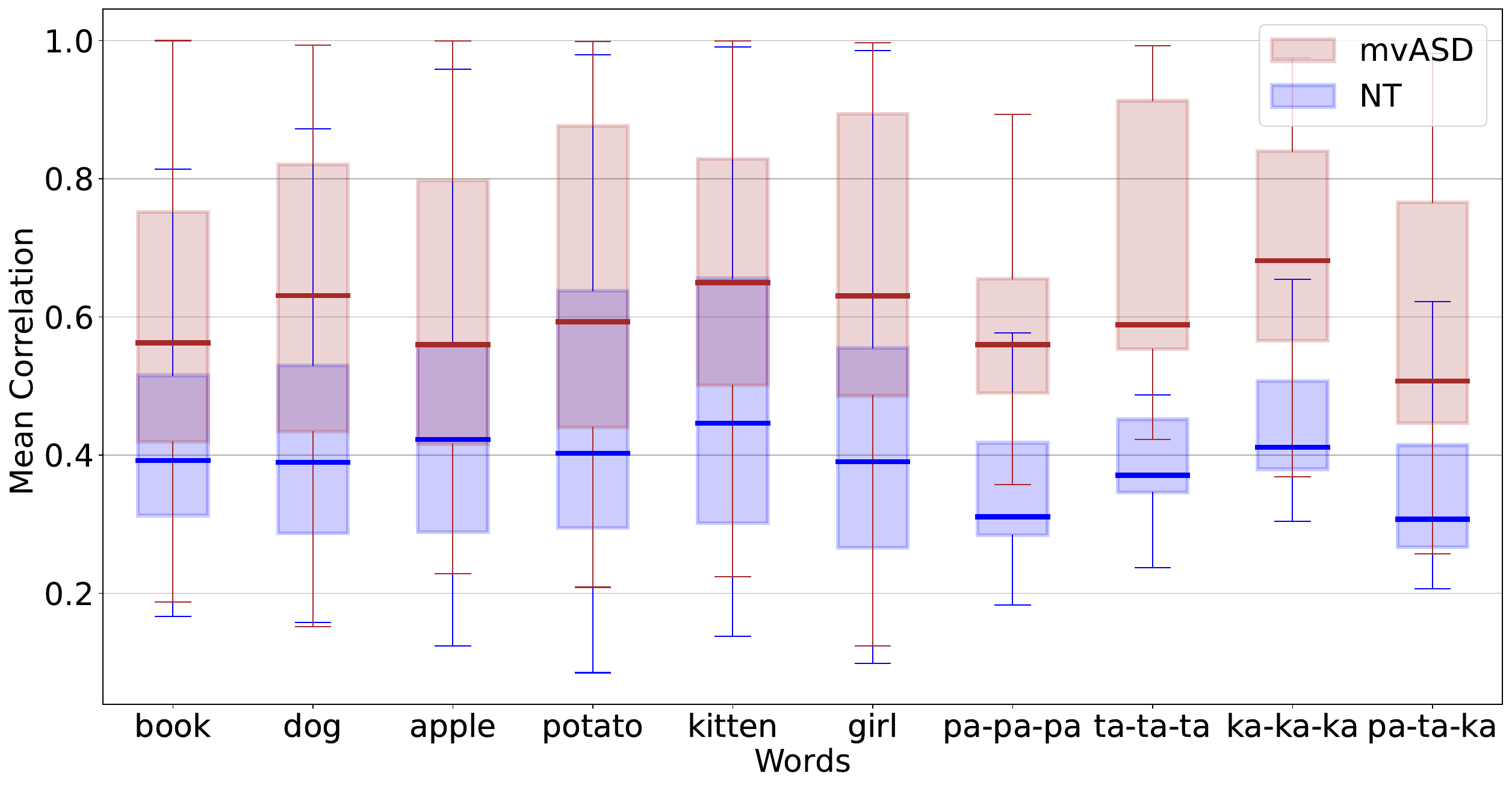}}
    
    
    \caption{Correlation between the sEMG channels}
    \vspace{-3mm}
\end{figure}

\subsection{Correlation of Facial Muscles}
Figure \ref{fig:correlation_matrix} presents the cross-correlation matrix and Figure \ref{fig:correlation} illustrates box plots comparing the average correlation between the different sEMG channels for each utterance in the mvASD and NT groups. We observe that the mvASD group has higher average correlation values compared to the NT group for all the words and DDK sequences ($p=0.012$). The p-values as shown in Table \ref{tab:features}, reveal that the individual word/sequence differences were also significantly greater for the mvASD group. The higher mean correlation in the mvASD group suggests greater synchrony in neuromuscular action potentials and tighter coupling among the different muscle groups.


\subsection{Complexity of Motor Coordination}

\begin{figure}[!h]
    \subfloat[Diadochokinetic (DDK) sequencing task.]{\label{fig:cohen_ddk}
    \includegraphics[width=0.9\linewidth]{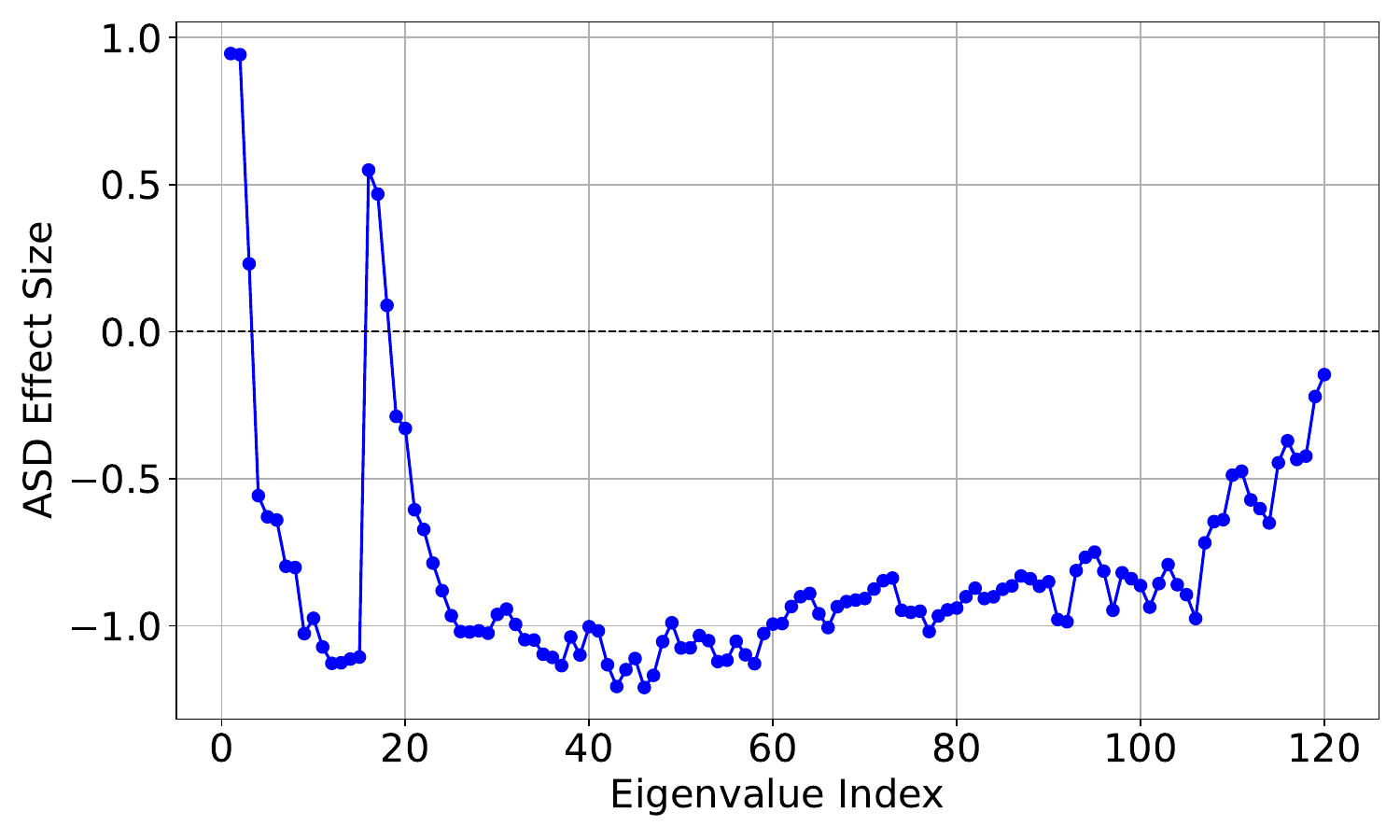}}
    
    \subfloat[Word production task.]{
    \label{fig:cohen_word}
    \includegraphics[width=0.9\linewidth]{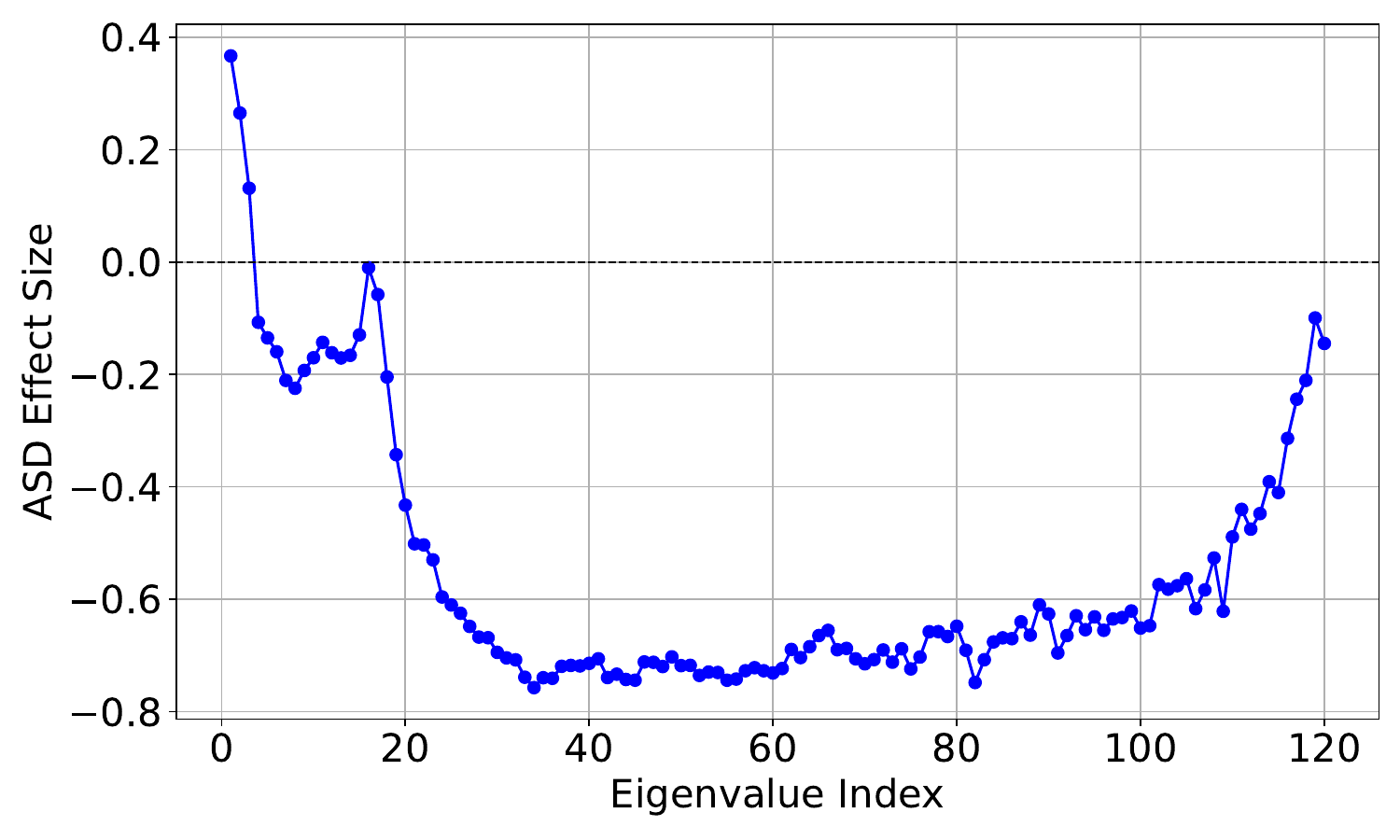}}
    \caption{Cohen D effect sizes for eigenvalues derived from auto- and cross-correlation matrices during speech tasks.}
    \vspace{-3mm}
\end{figure}

Given that a greater correlation between the activity of the facial muscles involved during the speech tasks in mvASD compared to NT may imply a stronger interdependence of these muscles, we compute Cohen's d effect sizes of the eigenvalue features derived from the time-delayed auto- and cross-correlation matrices for the DDK task and word task, as shown in Figure \ref{fig:cohen_ddk} and Figure \ref{fig:cohen_word} respectively. An effect size higher than $0$ indicates greater eigenvalues in the mvASD group compared to the NT group. This demonstrates that the mvASD participants have lower complexity in the underlying signals of their muscle movement patterns compared to neurotypical NT participants.

\section{Discussion and Future Work}

This paper presents the measurement and analysis of surface electromyography (sEMG) signals obtained from eight facial muscles associated with speech during diadochokinetic sequencing and word production tasks, in mvASD adults and neurotypical adults. The analysis showed a higher mean RMS and significantly greater mean correlation between the individual sEMG channels in the mvASD group compared to the neurotypical (NT) group. This suggests higher muscle activation and synchrony in the motor unit discharge pattern from the facial muscles during speech production, highlighting previous evidence of speech challenges linked to impairments in oromotor skills within the mvASD population \cite{maffei2023oromotor}, but by utilizing a more direct assessment of muscle activity through physiological sEMG signals. Moreover, we find that individuals with mvASD exhibit reduced complexity in muscle coordination, evident by a rapid decline in eigenvalues derived from correlation matrices, unlike the NT group which shows higher complexity as demonstrated by a more uniform distribution of eigenvalues. However, this method is constrained by possible signal noise, reduced high-frequency content, and phase shifts potentially influencing muscle complexity. Further investigation into the
contribution of different muscles to the differences in complexity would be useful. The neural drive may also be underestimated by sEMG as shown by Farina et al \cite{farina2004extraction}.

The use of only a small set of words while not optimal is in keeping with the limited speech capabilities in this population. The results presented here pave the way for new hypotheses and analyses of speech in mvASD. Future examination of sEMG signals in relationship to the phonemic representations of the speech targets, characterizing the preparatory phase preceding speech initiation, and exploring correlations of standardized assessments with multimodal data from audio, video,
and handwriting data present promising avenues for additional research.




\section{Acknowledgements}
This work was supported by funding from the Nancy Lurie Marks Family Foundation.
\bibliographystyle{IEEEtran}
\bibliography{mybib}

\end{document}